\documentclass[a4paper,fleqn,usenatbib,useAMS,referee]{mnrasBSAO}
\usepackage[normalem]{ulem}
\usepackage{graphicx}	
\usepackage{amsmath}	
\usepackage{amssymb}	
\usepackage{multicol}        
\usepackage{bm}		
\usepackage{pdflscape}	
\usepackage{epstopdf}
\usepackage{hyperref}

\usepackage[T1]{fontenc}
\usepackage{ae,aecompl}

\usepackage{txfonts}

\title[LRN 2015 in M\,101]{Luminous Red Nova 2015 in the Galaxy M\,101}

\author[V. P. Goranskij et al.]{
V.~P.~Goranskij,$^{1}$
\thanks{Contact e-mail: \href{mailto:goray@sai.msu.ru}{goray@sai.msu.ru}}
~E.~A.~Barsukova,$^{2}$
~O.~I.~Spiridonova,$^{2}$
~A.~F.~Valeev,$^{2,3}$
\newauthor 
~T.~A.~Fatkhullin,$^{2}$
~A.~S.~Moskvitin,$^{2}$
~O.~V.~Vozyakova,$^{1}$
~D.~V.~Cheryasov,$^{1}$
\newauthor 
~B.~S.~Safonov,$^{1}$
~A.~V.~Zharova,$^{1}$
~T.~Hancock$^{4}$
\\
 $^{1}${Sternberg Astronomical Institute, Moscow University, Universitetsky Prospect, 13, Moscow 119899 Russia}\\
 $^{2}${Special Astrophysical Observatory of Russian Academy of Sciences, Nizhnij Arkhyz, Karachay-Cherkessia  369167 Russia}\\
 $^{3}${Kazan Federal University, ul. Kremlevskaya 18, Kazan, Tatarstan 420008 Russia}\\
 $^{4}${Downunder Observatory, Fremont, MI, USA}
}

\date{Manuscript November, 2016}

\pubyear{2016}

\begin{document}
\label{firstpage}
\pagerange{\pageref{firstpage}--\pageref{lastpage}}
\maketitle

\vspace{-3mm}
\begin{abstract}
We present the results of the study of the red nova PSN\,J14021678
+5426205 based on the
observations carried out with the Russian 6-m telescope (BTA) along
with other telescopes of SAO RAS and SAI MSU. To investigate the nova
progenitor, we used the data  from the Digital Sky Survey and amateur
photos available on the internet. In the period between April 1993
and July 2014, the brightness of the progenitor gradually increased
by $2\fm2$ in the $V$ band. At the peak of the first outburst in
mid-November of 2014, the star reached an absolute visual magnitude of
$-12\fm75$ but was discovered later, in February 2015, in a repeated
outburst at the absolute magnitude of $-11\fm65$. The amplitude of the
outburst was minimum among the red novae, only $5\fm6$ in the
$V$ band. The H$\alpha$ emission line and the continuum of a cool
supergiant with a gradually decreasing surface temperature
were observed in the spectra. Such process is typical for red novae,
although the object under study showed extreme parameters: maximum
luminosity, maximum outburst duration, minimum outburst amplitude,
unusual shape of the light curve. This event is interpreted as a
massive OB star system components' merging accompanied by the formation
of a common envelope and then the expansion of this envelope with
minimal energy losses.
\end{abstract}

\vspace{-4mm}


\section{INTRODUCTION}

Luminous red novae are representatives of a sparsely populated class
of exploding variables which is known since 1988 when such a star
exploded in the M31 galaxy. This was Red Variable = McD\,88 No.1 =
M31\,V1006/7 \citep{Rich89:Goranskij_n,Shar93:Goranskij_n}. Among the
Galactic red novae, V4332\,Sgr, V838 Mon, and V1309\,Sco are the most
thoroughly studied. According to the archival data, the Galactic novae
CK\,Vul = N\,Vul\,1670 \citep{Hayd1:Goranskij_n,Kamin15:Goranskij_n},
V1148 Sgr = N\,Sgr\,1943 \citep{May49:Goranskij_n}, and
OGLE-2002-BLG-360 \citep{Tyl13:Goranskij_n} are supposed to belong to
this class. The most precise phenomenological definition for the
stars of this type is given by~\cite{Munari02:Goranskij_n}~---Stars
Erupting into Cool Supergiants (SECS). Such ``cool explosions'' were
not predicted theoretically.  The absolute magnitudes of red novae
exceed those of classical novae in maximum brightness reaching
$M_V = -12^{\rm m}$ and $M_R = -12\fm3$ (OT\,2006-1 in M\,85,
\citet{Kul07:Goranskij_n,Pastor07:Goranskij_n} , and PTF10fqs in M\,99,
\citet{Kasli11:Goranskij_n}) and fall into the interval of magnitudes
between classical novae and supernovae ($-17^{\rm m} < M_V < -8^{\rm m}$,
\citet{Berger09:Goranskij_n}). Red novae together with other
optical transients which magnitudes are within this interval are
called Supernova (SN) Impostors, or Intermediate Luminosity Red
Transients (ILRT).

When the majority of researchers explain the physical nature of red
novae, they prefer a merger of components in a binary or multiple
system \citep{SoTy03:Goranskij_n,SoTy06:Goranskij_n} and call them
``mergers''. The observations of the star V1309\,Sco using the OGLE
archive data confirmed this hypothesis~\citep{Tylen11:Goranskij_n}.
Six years prior to the outburst, this star was a \mbox{W\,UMa}-type
contact binary with an orbital period of 1.44~days. The process of
merging of the components that ended with the red nova outburst was
directly observed in this system. It is supposed by
\cite{Bars14:Goranskij_n} that the red nova phenomenon is
associated with the energy burst in the stellar core after which the
envelope passes into expansion stage similar to the adiabatic one
(with minimum energy loss); the nova outburst occurs with a delay of
a year or even several years. In order to explain the phenomenon of
the red nova V4332\,Sgr, the hypothesis of a ``slow shock'' which
forces a stellar photosphere to expand has been suggested by
\cite{Martini99:Goranskij_n}. This shock can be caused both by
merging of the stellar cores of a binary system and by an
instability of the core of a single star. V838\,Mon is an example of
a red nova not associated with the component
merging~\citep{Goray14:Goranskij_n}. It is a wide detached system
containing the B3\,V-type component~\citep{Wagner03:Goranskij_n} which did
not participate in the outburst of 2002 but was later engulfed by the
explosion remnant. Thus, there can be objects of different nature
among red novae. The remnants of some red novae contain dust and cool
rarified gas emitting in atomic and molecular lines.

The most valuable information on the red nova phenomenon can be
obtained  using the archival data and studying the stars, for which
the accurate distances are possible to determine. Such objects can be
situated in the nearby galaxies, however one usually needs large
telescopes to observe them.

In the first half of 2015, two red novae have been
discovered in the nearby galaxies: MASTER\,J004207.99+405501.1 in M\,31 =
M31N\,2015-01a~\citep{Shum15:Goranskij_n,Willi15:Goranskij_n,Kurten15:Goranskij_n}
and PSN\,J14021678 +5426205 or Luminous Red Nova, LRN\,2015 in M\,101
\citep{Gerke15:Goranskij_n,Cao15:Goranskij_n,Vinko15:Goranskij_n,Kelly15:Goranskij_n,Goray15:Goranskij_n}.
M31N\,2015-01a is similar to  V1006/7 in M\,31 at the brightness
maximum  \mbox{$V = 15\fm4$} and with the absolute magnitude
\mbox{$M_V=-9^{\rm m}$}, although, its outburst duration was twice
shorter than that of V1006/7. The  LRN\,2015 in M\,101 reached the absolute
magnitude at the maximum of $M_V = -12\fm75$, and demonstrated
unusual properties not observed earlier in other red novae. It is a
SN impostor in its absolute magnitude at the brightness maximum. Four
actual SNe of I and II types were observed earlier in M\,101: 1909A,
1951H, 1970G, and 2011fe. The present paper is dedicated to the
investigation of the LRN in M\,101.

The luminous red nova PSN\,J14021678+5426205 in M\,101 was discovered by
C.~D.~V{\^\i}ntdevar\v{a}\footnote{{\tt
http://www.rochesterastronomy.org/snimages/}} at the B{\^a}rlad
Astronomical Observatory in Romania on February 10, 2015. According
to our measurements of the CCD frame in which it was discovered, the
star brightness was $17\fm50$ in~$V$ band. According
to~\cite{Cao15:Goranskij_n}, on November 10, 2014 the star was
brighter, $16\fm36$, in the $R$ filter. However, the observations on
January 19, 2015 by~\cite{Gerke15:Goranskij_n} showed the star was
considerably fainter, $R=18\fm23$ and $V=18\fm80$. These observations
proved that the brighter outburst took place in November 2014, after
which the star brightness notably weakened. The red nova was
discovered in the repeated outburst, the brightness maximum of which
was in February 2015.

\section{STUDY OF THE NOVA PROGENITOR WITH ARCHIVAL DATA}

The earliest published SDSS sky survey observations of the progenitor
of the LRN outburst in  M\,101 refer to March 2003. Stellar
magnitudes in the $ugriz$ system relative to Vega in these filters
are \mbox{$21.1\pm0.3$}, \mbox{$21.6\pm0.3$}, \mbox{$21.0\pm0.3$},
\mbox{$20.6\pm0.3$}, and \mbox{$21.9\pm0.9$}
respectively~\citep{Kelly15:Goranskij_n}. One can determine the value
\mbox{$V=21\fm2$} and color indices $B-V=0\fm4$ and \mbox{$V-R_c =0\fm3$}
in the Johnson--Cousins system using the interpolation
method. According to the archival data of the Large Binocular
Telescope (LBT) from the middle of 2012 to the middle of 2014 the
brightness of the nova progenitor star increased from  $20\fm97$ to
$19\fm78$ in the $V$ band, and from $20\fm69$ to $19\fm59$ in the
$R$ band \citep{Gerke15:Goranskij_n}.

\begin{figure}

\vspace{2mm}
\begin{center}
\includegraphics[width=0.8\columnwidth]{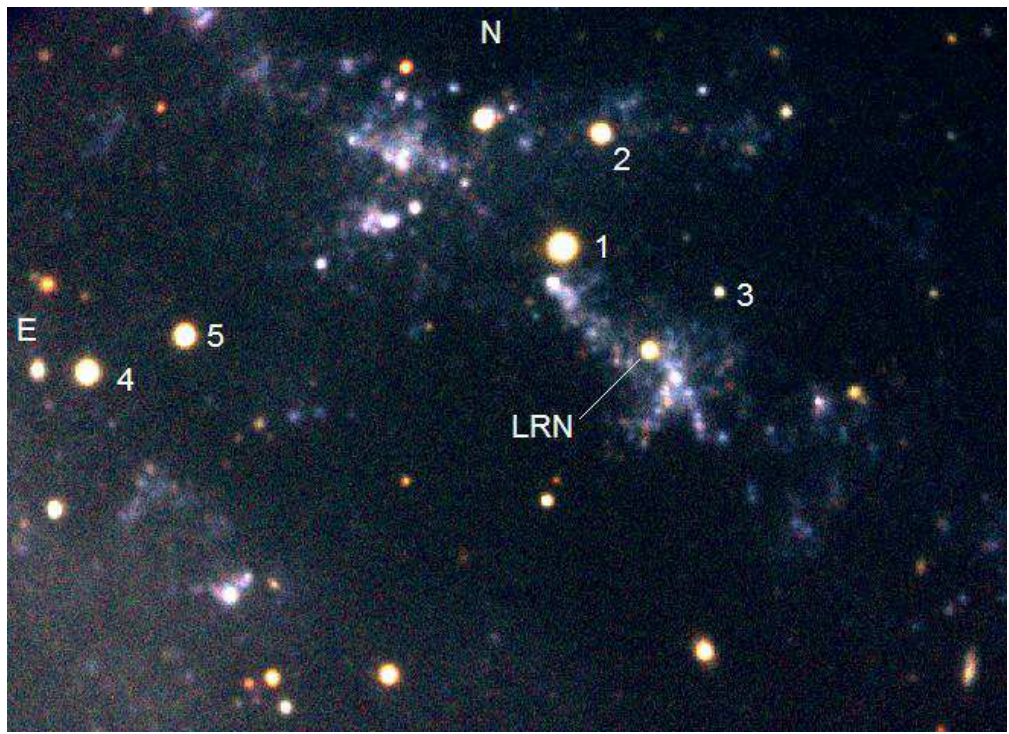}
\end{center}
\caption{Map of the LRN in M\,101 and standard stars, February
24, 2015 (BTA/SCORPIO). We show the region of the $4\farcm9\times3\farcm6$ size.
The electronic image is in color.}
\label{figure1:Goranskij_n}
\end{figure}

We found only one faint image of the star in the DSS sky survey
archives on the photo from the Palomar sky survey POSS-II taken with
the Palomar 48-inch Schmidt telescope on April 15, 1993 using the
Kodak\,IIIaJ emulsion. The maximum sensibility of this emulsion lies
between the $B$ and $V$ bands. We estimate the brightness of the
star at that moment as $22\fm0\pm0.3$ in $V$ band. To conduct the
photometric measurements we created a local standard in the vicinity
of the LRN in M\,101 with a reference to the standard near the blazar
S4\,0954+65 \citep{Rai99:Goranskij_n}, and then we extended it to weak
magnitudes. The map in Fig.~\ref{figure1:Goranskij_n} shows our
standard stars and Table~\ref{table1:Goranskij_n} contains their
coordinates and values.

\begin{table}
\begin{center}
 
\caption{Coordinates and $BVR$ magnitudes of the comparison stars in the vicinity of the LRN in M\,101}\label{table1:Goranskij_n}

\begin{tabular}{c|c|c|c|c|c}

\hline
Star & RA,    &   Dec,  &   $B$,  &   $V$,  &  $R_c$, \\
No.  &  hh mm ss   &   dd mm ss   &   mag &   mag &  mag  \\
\hline
1 & 14 02 19.12 & 54 26 57.2 &\ 16.044 &\ 15.030 &\ 14.276 \\
  &           &          & $\pm$0.019 & $\pm$0.018 & $\pm$0.019 \\
2 & 14 02 16.94 & 54 27 29.2 &\ 18.006 &\ 16.879 &\ 16.024 \\
  &           &          &\ $\pm$0.02  & \ $\pm$0.02  &\ $\pm$0.02 \\
3 & 14 02 13.94 & 54 26 34.2 &\ 20.135 &\ 19.555  &\ 19.158  \\
  &           &          & $\pm$0.033 & $\pm$0.034 & $\pm$0.055 \\
4 & 14 02 36.42 & 54 26 46.9 &\ 16.996 &\ 16.100 &\ 15.476 \\
  &           &          & $\pm$0.018 & $\pm$0.018 & $\pm$0.045 \\
5 & 14 02 32.88 & 54 26 52.0 &\ 17.496 &\ 16.460 &\ 15.684 \\
  &           &          & $\pm$0.014 & $\pm$0.012 & $\pm$0.039 \\
\hline

\end{tabular}
\end{center}
\end{table}

One can find the image of the LRN progenitor in many amateur color
photos of the galaxy M\,101 on the internet. Note that they contain
color information (see {\tt Flickr.com}, for example). So, the blue
color of the star is seen in the images of \mbox{2011--2013}. One of
the authors of this paper, Terry Hancock, carried out his
observations using the Astro-Tech 250-mm Ritchey--Chr\'etien
astrograph with the monochrome camera QHY9M (with the Kodak
chip~KAF\,8300) and RGB-filters. The exposures were accumulated
during 24 hours in five nights between March 14 and 27, 2012
(JD~$\approx$~2456009). Our photometry of these images in the FITS
format with reference to the standard in the $BVR_c$ system gives the
following magnitudes and accuracy estimates:
$$
B = 21.34^{-0.19}_{+0.22};~ V = 21.06^{-0.15}_{+0.16};~  R_c =
20.63^{-0.20}_{+0.25}.
$$
We converted color images in other formats into the BITMAP format and
measured separate RGB components with the reference to the standard
$R_c$, $V$, and~$B$ respectively. The accuracy of measurements kept
within the limits of $0\fm2$--$0\fm3$. We also conducted the
photometry of the images by K.~Itagaki, the references to which were
given in the Central Bureau for Astronomical Telegrams
(CBAT)\footnote{{\tt
http://www.cbat.eps.harvard.edu/unconf/tocp.html}} of the
International Astronomical Union. The observations by Itagaki confirm
the first  and brightest star outburst in November 2014 and its
further weakening by $2\fm5$ prior to the second outburst. These
estimates of the brightness and color indices are in good agreement with
the SDSS and LBT observations.

\begin{figure}
 \vspace{2mm}
\begin{center}
\includegraphics[width=0.6\columnwidth]{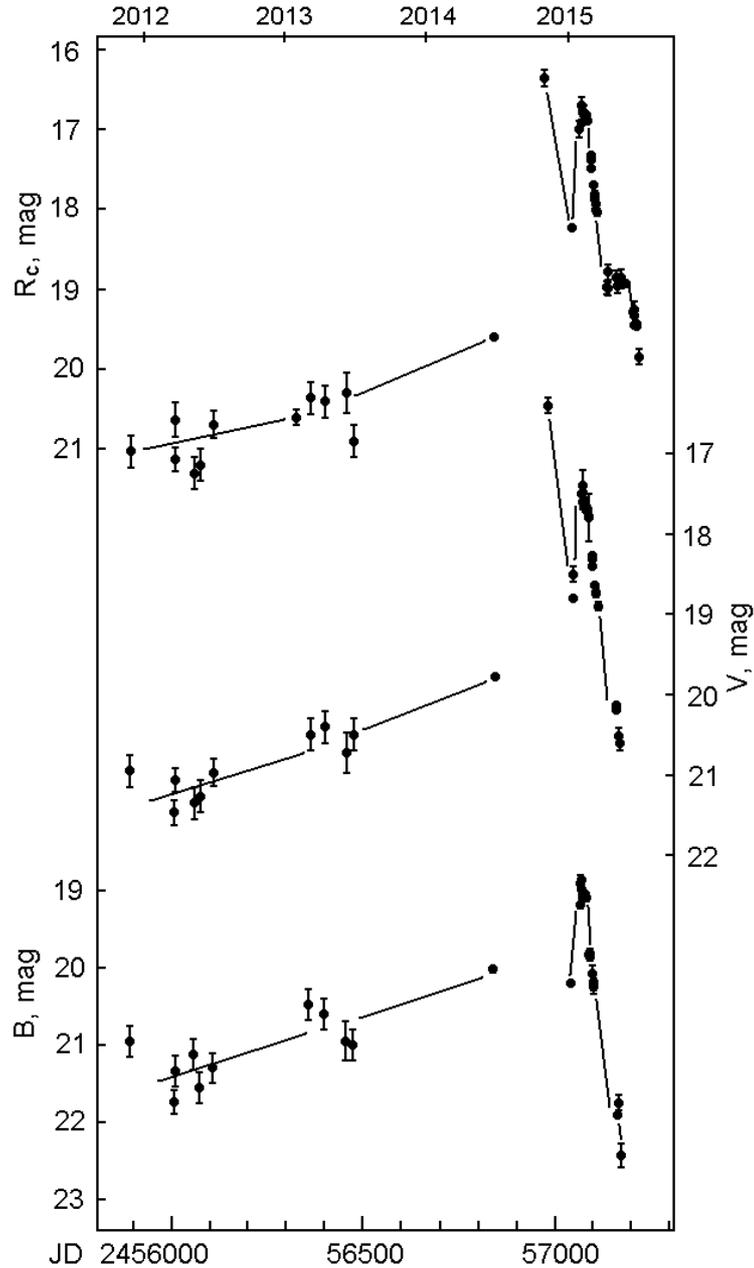}
\end{center}
\caption{Light curves of the LRN in M\,101,
2012--2015 in the $B$, $V$, and $R_c$ bands (bottom to top). }
   \label{figure2:Goranskij_n}
\end{figure}

\begin{figure}
 \vspace{5.5mm}
\begin{center}
\includegraphics[width=0.6\columnwidth, angle=0]{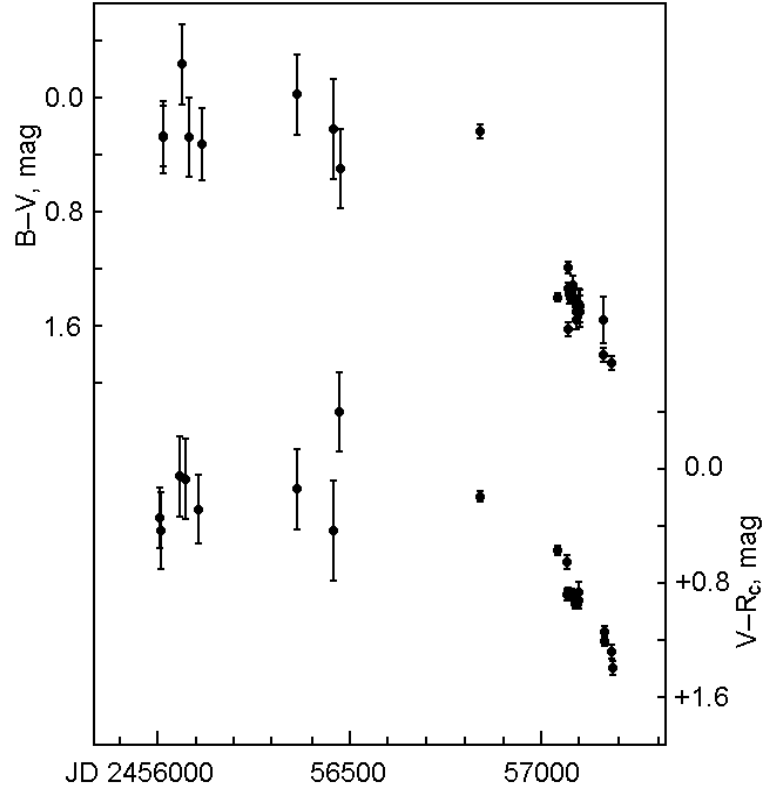}
\end{center}
\caption{Color-index curves $B - V$ (top) and
$V - R_c$ (bottom) constructed for the LRN in M\,101 using the
observations in 2012--2015.} \label{figure3:Goranskij_n}
\end{figure}

Table~\ref{table2:Goranskij_n} shows the results of our measurements
of the Digital Sky Survey (DSS) images, the CBAT, and the amateur
ones in the period from 1993 to 2015, before the discovery, and the
published data referring to this time period. The light curves in the
$BVR_c$  filters prior to and during the outburst, the color indices
$B-V$ and $V-R_c$ are presented in Fig.~\ref{figure2:Goranskij_n}
and \ref{figure3:Goranskij_n}. They may be studied in detail
interactively with a Java-compatible browser\footnote{{\tt
http://jet.sao.ru/$\sim$goray/psn1402.htm}}.

\begin{figure}
\begin{center}
\includegraphics[width=0.6\columnwidth]{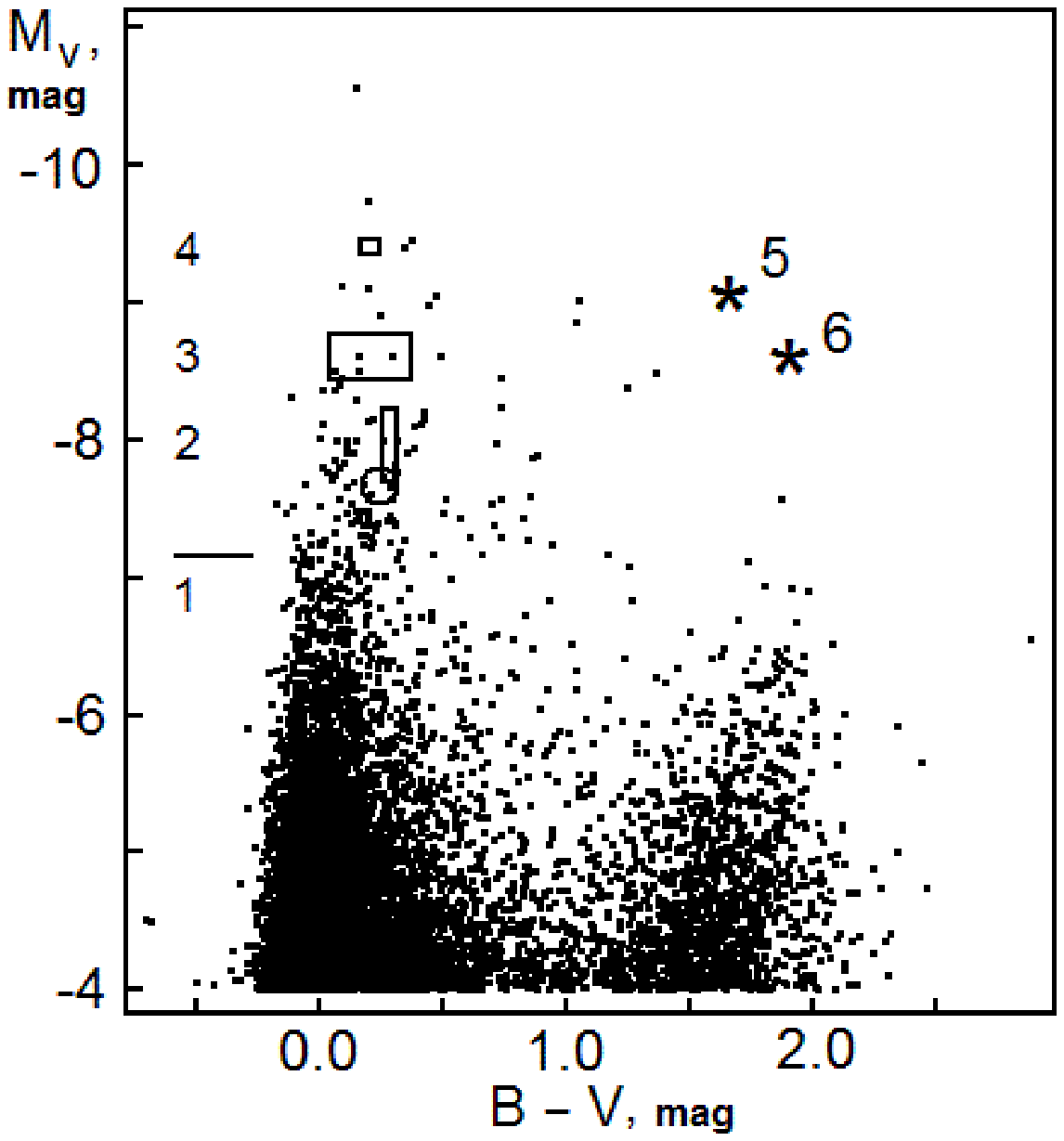}
\end{center}

\caption{Shift of the LRN progenitor in M\,101
in the ``color--absolute magnitude'' diagram before the outburst in
2014. (1)~Brightness  level of the star in 1993 according to the data
from POSS-II. The rectangles correspond to its position in the
diagram in 2012 (2); in 2013 (3); in the middle of
\mbox{2014}~(4)~{\protect \citep{Gerke15:Goranskij_n}}.
The asterisks denote the
position of the star at the stop of brightness decline in May 2015
(5) and in the middle of June 2015 (6). The circle denotes the
position of the massive semidetached system H$\alpha$19 from the
galaxy M\,33 with a high mass transfer rate in the phase of
components merging. The diagram was constructed for the 9492\_12
field in the galaxy M\,101.
} 
\label{figure4:Goranskij_n}

\end{figure}

These estimates show that the star brightness prior to the outburst
was gradually increasing from a level of $22\fm0$ in the~$V$ filter
(detected in 1993 in the DSS) up to $19\fm78$~$V$, stated with the
LBT in summer 2014. The data did not make it possible to determine if
there was a faster brightness variability during this brightness
increase. Deviations of some measurements from the
central trend do not exceed 3$\sigma$, as a rule. As it is known,
the orbital variability was observed in the V1309\,Sco red nova system
during the brightness increase prior to the outburst
\citep{Tylen11:Goranskij_n}. In the case of the LRN in M\,101,
there is unique information indicating that the pre-outburst
brightness increase happened with holding the surface temperature of
the star constant. The color indices stayed almost constant at
\mbox{$(B-V) \approx0\fm2$} and \mbox{$(V-R_c) \approx 0\fm2$}.
At that time, the star moved upward along
the main sequence of hot massive supergiants in the
``color--magnitude'' diagram $V-(B-V)$ (see
Fig.~\ref{figure4:Goranskij_n}). The diagram for normal stars in this
figure is plotted based on the photometry by
\cite{Grammer13:Goranskij_n} conducted with the Hubble Space
Telescope. The data are taken for the 9492\_12 field which is
situated closest to the site of outburst of the LRN in M\,101. The
red nova is located outside, in 36\arcsec\ from its eastern border.
The earliest measurements of LRN colors in 2012 show an evident
deviation of the star toward the red region of the main sequence
OB-supergiant branch; this can indicate the ending of the main
sequence stage and the beginning of evolution into red giant of the
brighter and more massive component of the system. In 2012, the
position of the star in the diagram was close to the position of the
well-known massive eclipsing system of high luminosity H$\alpha$19 in
the galaxy M\,33 denoted with the circle in
Fig.~\ref{figure4:Goranskij_n}. The orbital period of H$\alpha$19 is
equal to $33\fd108$, the absolute magnitude is $M_V = -7\fm6$. Mass
estimations of the H$\alpha$19 system components are
$40$--$50~M_\odot$ \citep{Goray10:Goranskij_n}. This is a semidetached
system with so high mass transfer rate that on the surface of its hot
component there is a bright spot associated with the circulation in
the envelope of this component of a gas from an accretion flow of
its companion and with the transfer of material from the depth of the
envelope to the surface. The contribution of this spot can be seen in
the light curve. The earliest observations of the LRN in M\,101 in
\hbox{1993} detected the star at \mbox{$M_V = -7\fm1$.} It is weaker
than H$\alpha$19 by $0\fm5$ only. Probably, the progenitor of the LRN
in M~101 was a similar massive system with the mass a bit smaller
than that of the H$\alpha$19 components.
\begin{table*}
\caption{Pre-discovery observations of the LRN/M101}\label{table2:Goranskij_n}
\begin{tabular}{l|c|c|c|c|l}
\hline
\multicolumn{1}{c|}{Date}         &JD\,2400000\,+& $B$, mag   &   $V$, mag  &  $R$, mag  &  \multicolumn{1}{c}{Source} \\
\hline
Apr 15, 1993    & 49093 &  --    &  22.0  &  --    & POSS II, Kodak IIIaJ       \\
Mar 07--10, 2003& 52707 & 21.6  &  21.2  & 20.90 & SDSS, ATel 7082$^{(1)}$       \\
Nov 25, 2011    & 55891 & 20.95 &  20.95 & 21.03 & R. Pecce, {\tt Flickr.com}       \\
Mar 20, 2012    & 56007 & 21.74 &  21.47 & 21.13 & D. Hartmann, Astrobin$^{(2)}$ \\
Feb 14--27, 2012& 56009 & 21.34 &  21.06 & 20.63 & T. Hankock, RGB images$^{(3)}$\\
May 10, 2012    & 56058 & 21.12 &  21.35 & 21.30 & O. Bryzgalov, {\tt Flickr.com}   \\
May 26, 2012$^*$)& 56074 & 21.55 &  21.27 & 21.20 & O. Bryzgalov, {\tt Flickr.com}   \\
Jan--Jun 2012   & 56109 & 21.30 &  20.97 & 20.69 & ATel 7069, LBT             \\
Feb 01, 2013    & 56324 &  --    &   --   & 20.60 & ATel 7070, PTF             \\
Apr 2013        & 56360 & 20.48 &  20.50 & 20.36 & Z. Orbanic, {\tt Flickr.com}$^{(4)}$\\
Mar--May 2013   & 56398 & 20.6  &  20.4  & 20.40 & R. Pfile, {\tt Flickr.com}       \\
Jun 11, 2013    & 56455 & 20.95 &  20.73 & 20.30 & S. Furlong, {\tt Flickr.com}     \\
Jun 29, 2013    & 56473 & 21.0  &  20.5  & 20.9  & C. Frenzi, {\tt Flickr.com}      \\
Jun--Jul 2014   & 56839 & 20.02 &  19.78 & 19.59 & ATel 7069, LBT             \\
Nov 10, 2014    & 56971 &  --    &   --    & 16.36 & ATel 7070, PTF             \\
Nov 13, 2014    & 56975 &  --    &  16.40 &  --    & K. Itagaki, CBAT$^{(5)}$      \\
Jan 19, 2015    & 57042 & 20.20 &  18.80 & 18.23 & ATel 7069, LBT             \\
Jan 20, 2015    & 57043 &  --    &  18.50 &  --    & K. Itagaki, CBAT$^{(5)}$      \\
Feb 10, 2015    & 57064.4 &   -- &  17.50 &  --    & C.~D. Vintdevara, discovery \\
 \hline
\multicolumn{6}{p{0mm}}{ 
\begin{tabular}{rp{123mm}}
 $^{(1)}$ & the SDSS-values in the $ugriz$ system (Vega) recounted to the $BVR_c$ system.\\
 $^{(2)}$ & {\tt http://www.astrobin.com/users/DetlefHartmann/} \\
 $^{(3)}$ & the 25-cm Ritchey-Chretien astrograph the CCD QHY9M Monochrome (Kodak KAF 8300 chip), 24-hour exposure. \\
 $^{(4)}$ & Uploaded on June 30, 2014. Approximate time of imaging, April 2013,
can be determined from the brightness of SN 2011fe. \\
 $^{(5)}$ & {\tt http://www.cbat.eps.harvard.edu/unconf/followups/J14021678+5426205.html.}\\
 $^*$) & Misprint in Astrophys. Bull. V.71, 82. This date is corrected.\\
\end{tabular}
 }
\end{tabular}
\end{table*}

\section{PRESENT-DAY PHOTOMETRY}

The photometric observations of the LRN in M\,101 in the repeated
outburst were carried out with several telescopes of SAO~RAS and
SAI~MSU in the $BVR_c$ system from February 15 to June 13, 2015.
Table~\ref{table3:Goranskij_n} shows the  photometry results, the
data on telescopes and instruments used are given in the last column.
We made the photometric reference to the comparison stars (see
Table~\ref{table1:Goranskij_n}). The mean measurement accuracy at the
level of \mbox{$16$--$18^{\rm m}$} was $0\fm01$, at the level of
\mbox{$19$--$20^{\rm m}$} it was about \mbox{$0\fm02$--$0\fm04$}, but at
the level weaker than $21^{\rm m}$ it could be up to~$0\fm1$.

\begin{table*}
\caption{The photometry of the LRN in M\,101}\label{table3:Goranskij_n}
\begin{tabular}{c|c|c|c|c||c|c|c|c|c}
\hline
JD$_{\odot}$\,2400000\,+&  $B$  &  $V$   &   $R_c$ & Remarks& JD$_{\odot}$\,2400000\,+&  $B$  &  $V$   &   $R_c$ & Remarks\\
\hline
57069.3642 &   --    & 17.664 & 16.776 & KG &     57105.5199 &   --    &   --    & 18.009 & SO \\
57069.4892 & 18.862 & 17.666 & 16.784 & KG &     57106.3993 &   --    & 18.901 & 18.041 & SO \\
57071.5603 & 18.988 & 17.649 & 16.771 & SO &     57133.2957 &   --    &   --    & 18.977 & SO \\
57071.5784 &   --    & 17.655 & 16.776 & SO &     57135.5288 &   --    &   --    & 18.991 & SO \\
57072.5776 & 19.034 & 17.667 & 16.803 & SO &     57136.4051 &   --    &   --    & 18.785 & SO \\
57072.5894 &   --    & 17.663 & 16.793 & SO &     57158.3471 &   --    &   --    & 18.859 & SO \\
57074.5439 & 19.086 & 17.708 & 16.825 & KG &     57160.3074 &   --    &   --    & 18.952 & SO \\
57075.5213 & 19.045 & 17.704 & 16.815 & SO &     57161.3043 &   --    &   --    & 18.915 & SO \\
57075.5326 &   --    & 17.669 & 16.808 & SO &     57162.3288 &   --    &   --    & 18.853 & SO \\
57076.5682 & 19.082 & 17.687 & 16.815 & SO &     57163.3073 &   --    &   --    & 18.928 & SO \\
57076.5790 &   --    & 17.690 & 16.808 & SO &     57164.3126 & 21.87  & 20.075 & 18.928 & SO \\
57077.5862 & 19.066 & 17.690 & 16.826 & SO &     57165.3147 & 21.70  & 20.160 & 18.954 & SO \\
57077.5793 &   --    & 17.699 &   --    & SO &     57166.3898 &   --    & 20.158 & 18.967 & SO \\
57078.3967 & 19.088 & 17.698 & 16.827 & SO &     57182.3213 &   --    &   --    & 19.291 & SO \\
57078.4095 &   --    & 17.708 & 16.829 & SO &     57183.2694 &   --    &   --    & 19.334 & CR \\
57078.4318 & 19.037 & 17.691 & 16.816 & 6m &     57183.3529 &   --    &   --    & 19.233 & SO \\
57081.5474 & 19.086 & 17.784 & 16.901 & KG &     57184.34   &   --    & 20.72  & 19.267 & SO \\
57081.5639 &   --    & 17.796 & 16.896 & KG &     57185.29   & 22.38  & 20.55  & 19.278 & SO \\
57090.4413 & 19.832 & 18.268 & 17.325 & SO &     57185.3360 &   --    &   --    & 19.352 & 6m \\
57090.4628 &   --    & 18.295 & 17.322 & SO &     57185.3373 &   --    &   --    & 19.313 & 6m \\
57091.5366 & 19.814 & 18.305 & 17.375 & SO &     57185.3382 &   --    &   --    & 19.316 & 6m \\
57091.5485 &   --    & 18.307 & 17.380 & SO &     57186.2240 &   --    &   --    & 19.332 & CR \\
57092.5591 & 19.856 & 18.399 & 17.489 & SO &     57186.3192 &   --    &   --    & 19.330 & SO \\
57097.3654 & 20.070 & 18.633 & 17.696 & SO &     57187.2308 &   --    &   --    & 19.374 & CR \\
57100.3466 & 20.190 & 18.730 & 17.814 & SO &     57187.3792 &   --    &   --    & 19.365 & SO \\
57100.3919 & 20.241 & 18.742 & 17.881 & SO &     57211.4180 &   --    &   --    & 19.820 & SO \\
57104.5638 &   --    &   --    & 17.931 & SO &                &        &        &        &    \\
\hline
\multicolumn{6}{p{0mm}}{ 
\footnotesize 
\begin{tabular}{lp{137mm}}
 6m~---&  6-m BTA telescope and the SCORPIO focal reducer with $BVR_cI_c$ filters \citep{Afanas05:Goranskij_n}.\\
 KG~---&  2.5-m telescope of the SAI MSU Caucasian Mountain Observatory with the CCD-cameras Proline\,KAF\,39000 and NBI\,2k2k with $BVR_cI_c$ filters.\\
 SO~---& 1-m Zeiss telescope of SAO RAS and the $UBVR_cI_c$-photometer with the CCD EEV\,42-40.\\
 CR~---& 0.6-m Zeiss telescope of the Crimean laboratory of SAI MSU and the $UBVR_cR_jI_j$-photometer with the CCD camera Apogee-47p.\\
\end{tabular}
 }
\end{tabular}
\end{table*}

\begin{table*}
\caption{ Spectra of the LRN in M\,101 obtained with the
BTA/SCORPIO$^*$)}\label{table4:Goranskij_n}
\begin{tabular}{c|c|c|c|c|c|c|r}
\hline
Date & JD$_\odot$\,2400000\,+& $\epsilon$, s & $\lambda$, \AA & $R$, \AA & Grism & $\Delta v_r$, km\,s$^{-1}$ & $S/N$ \\
\hline
Feb 24, 2015 & 57078.4531 & 2400 & 4052--5848 & 5.0 & VPHG1200G & +5.0 & 180 \\
Feb 24, 2015 & 57078.5829 & 2751 & 5751--7498 & 5.0 & VPHG1200R & +4.9 &  50 \\
Jun 11, 2015 & 57185.3412 & 3600 & 4000--7919 & 14 & VPHG500G & $-$14.9  &  12 \\
\hline
\multicolumn{4}{l}
{\footnotesize $^*$) Misprint in Astrophys. Bull. V.71, 82. Dates are corrected.}
\end{tabular}
\end{table*}

According to the observations of the LRN in M\,101 carried out before
its discovery, during the first outburst the star reached $16\fm4$ in
the $V$ filter at the time JD\,2456975.3 and $16\fm36$ in the $R_c$
filter at the time JD\,2456971. This corresponds to the absolute
magnitudes $M_V \approx -12\fm75$ and \mbox{$M_R \approx -12\fm80$}.
As it is seen in the light curve (Fig.~\ref{figure2:Goranskij_n}),
the star brightness in the $V$ filter at the maximum of the first
outburst was higher by at least $1\fm1$ than at the maximum of the
second outburst. The same difference between the values in the
outbursts turned out to be significantly smaller in the $R_c$ filter,
$0\fm4$. From the observations in the first maximum, we cannot
estimate the stellar temperature in the luminosity peak, as the
response curves of the amateur instruments are not known exactly; it
is obvious, however, that the star was hotter in the peak of the
first outburst than in the peak of the second one. At the maximum of
the second outburst, the color indices of the star considerably
increased compared to those before the outburst at the end of the
gradual brightness increase in summer 2014. The $B-V$ color index
increased from $0\fm2$ to $1\fm3$, and the $V-R_c$ increased from
$0\fm2$ to $0\fm9$ (Fig.~\ref{figure3:Goranskij_n}). In the $R_c$
band, the star was observed more intense than in other bands and the
time of the secondary maximum can be reliably estimated as
JD\,2457069. At that, the color indices already were:
\mbox{$B-V=1\fm36\pm0.03$} and \mbox{$V-R_c=0\fm87\pm0.01$} and
corresponded to the spectral class \mbox{K2\,I--K3\,I}. The
brightness decrease in the $R_c$ filter in the period of
JD\,2457080--2457132 continued with an average velocity of $0\fm041$
per day; and then stopped for about $30$ days at the brightness level
of  $18\fm95\,R_c$. Later, the brightness decline continued with the
velocity twice lower. In the $V$ and $B$ filters, the brightness
decline rates were notably higher, $0\fm044$ and $0\fm055$ per day
respectively. The brightness decline stopping or slowing down took
place in these filters too, although, the data available do not allow
us to observe them so thoroughly as in the $R$ filter. In the
``color--magnitude'' diagram, the position of the star at the moment
of the brightness stop is marked with an asterisk and the number 5.
At that time, the star was a red supergiant with the luminosity
exceeding that one of the extreme red supergiant from the field
9492\_12 in the galaxy M\,101 \citep{Grammer13:Goranskij_n} by
$1\fm7$. After the brightness decrease stopped, the star went on
evolving toward red supergiants and in June 2015 reached the position
marked with an asterisk and the number 6 in
Fig.~\ref{figure4:Goranskij_n}. Multicolor photometry conducted on
June 11 yields the following values:  $V=20\fm55$; $B-V=1\fm83$;
$V-R_c=1\fm27$.

Based on the photometry results, the star can be reliably classified
as a luminous red nova, although, its light curve has a specific
feature distinguishing it from other red novae: the repeated
outburst. Repeated outbursts were also observed in the red nova
V838\,Mon, however, with a considerably smaller amplitude. Those
outbursts were explained by shock waves coming out to the surface,
which were caused by pulsations
\cite{Bars02:Goranskij_n,Goray07:Goranskij_n}, or swallowing of three
massive planets by an expanding red
giant~\cite{Retter03:Goranskij_n}.

\section{SPECTROSCOPY}

\begin{figure*}
 \vspace{2mm}
\includegraphics[scale=0.65, angle=0]{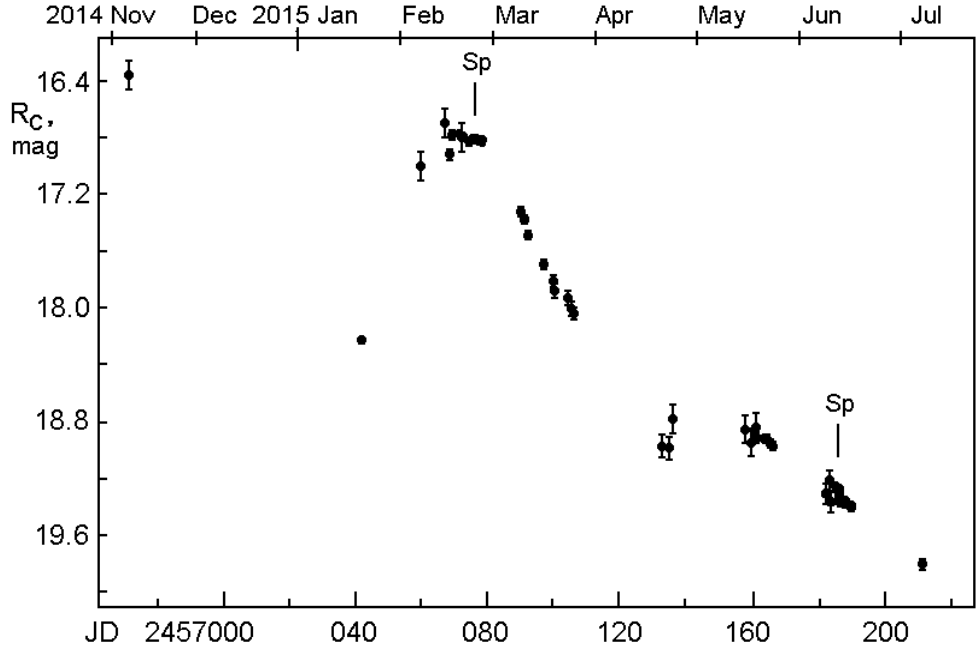}
\caption{Light curve of the LRN in M\,101 in
the $R_c$ band constructed for the second outburst. The times of
taking the BTA/SCORPIO spectra are noted as Sp.}
\label{figure5:Goranskij_n}
\end{figure*}

Spectroscopic observations of medium resolution were carried out at
SAO~RAS on the 6-m BTA telescope with the SCORPIO focal
reducer~\citep{Afanas05:Goranskij_n} on February 24 and June 11, 2015.
Table~\ref{table4:Goranskij_n} contains the main data on the obtained
spectra: date, Julian day, total exposure in seconds, spectral range,
spectral resolution, grism, heliocentric correction, and the
signal-to-noise ratio in the continuum in the middle of the spectral
range. The reduction of the spectra was conducted in the OS Linux
using the ESO~MIDAS medium and the LONG context (for the long-slit
spectra). The spectra obtained on June 11, 2015 were distorted by
fringes at wavelengths $\lambda >$\,6800~\AA. To compensate the
fringes, we took exposures with the shift of the star along the slit
and, as a result of subtracting the shifted spectra, the fringes were
removed. As the signal-to-noise ratio in the total spectrum was too
small, we smoothed the data using moving average method with an
averaging interval of 14~\AA, which was equal to the actual spectral
resolution. To convert the spectra into energy units, we used the
spectrophotometric standards HZ\,44 and GRW\,+70\degr5824 by
\cite{Oke90:Goranskij_n}, and simultaneous photometric observations.
In digitized form, the spectra are available on the
internet.\!\footnote{{\tt
http://jet.sao.ru/$\sim$bars/spectra/psn1402/}} In the $R_c$ light
curve (Fig.~\ref{figure5:Goranskij_n}), the times of the spectral
observations are denoted as follow: near the maximum of the second
outburst and at the brightness decline which followed after its stop.

\begin{figure*}
\includegraphics[scale=0.8]{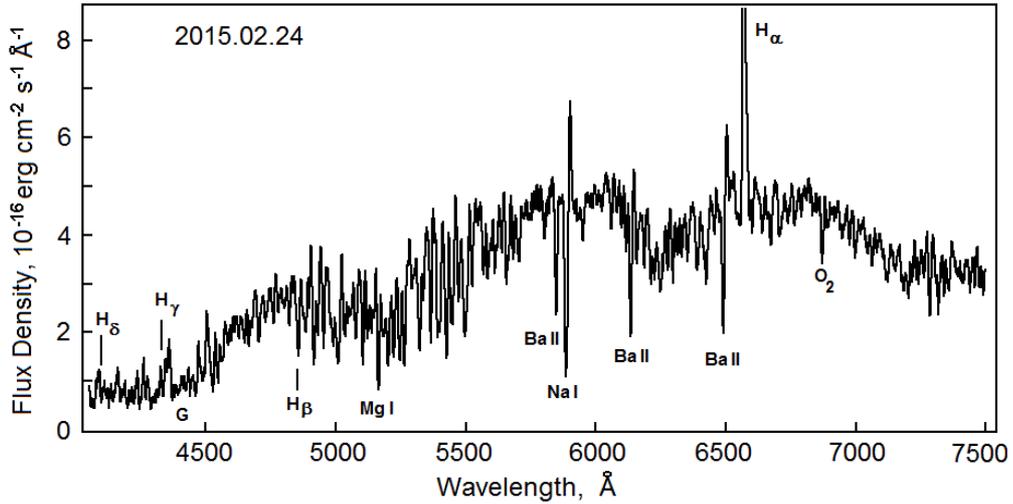}
\caption{Spectrum of the LRN in M\,101 obtained
on the BTA telescope with the SCORPIO focal reducer on February 24,
\mbox{2015} near the maximum of the second outburst. }
\label{figure6:Goranskij_n}
\end{figure*}

\begin{figure*}
\includegraphics[scale=0.7]{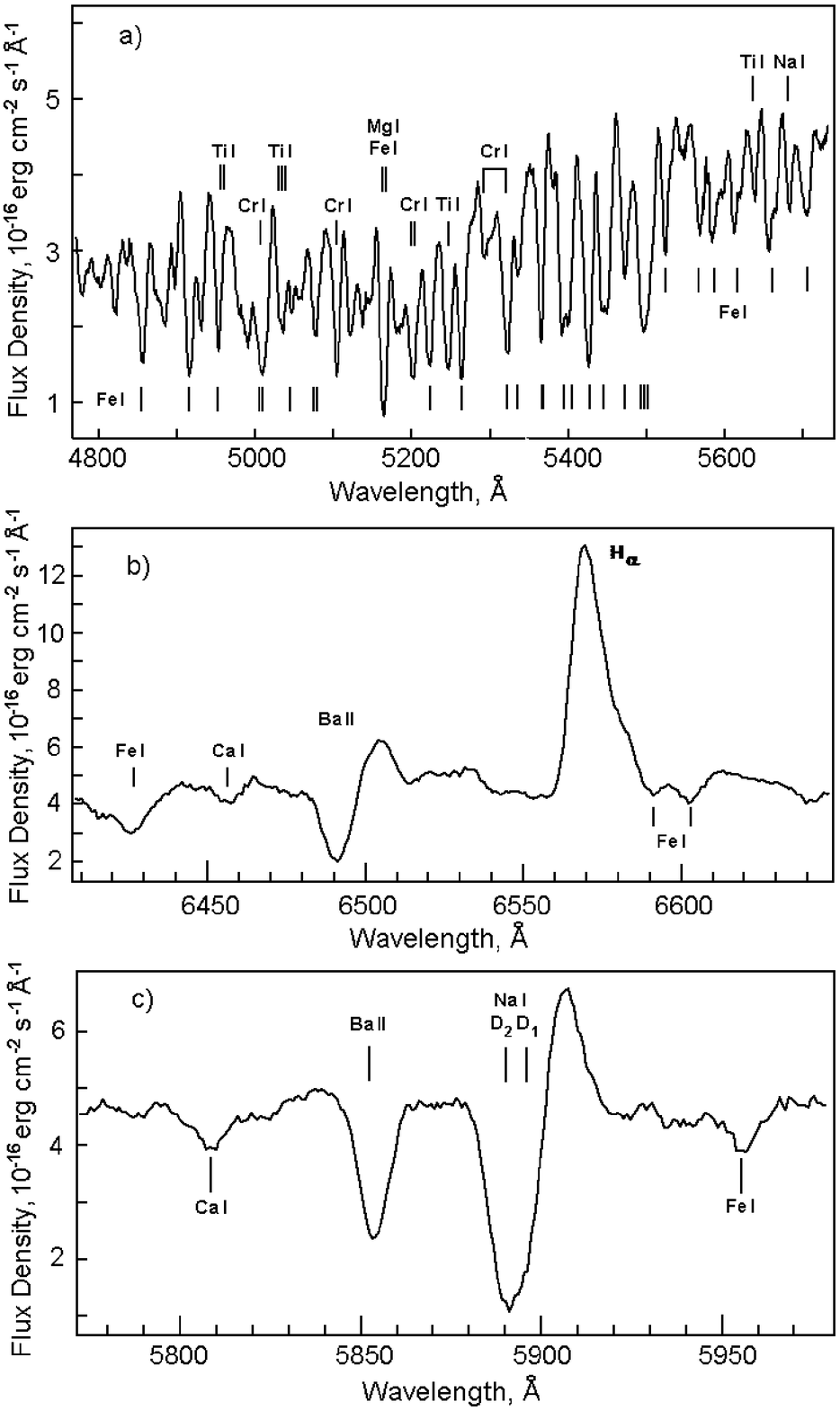}
\caption{Fragments of the spectrum of the LRN
in M\,101 obtained on February 24 and identification of the spectral
lines. (a) the depression region $\lambda$4800--5700~\AA\ associated
with the absorption of neutral atoms. (b)~the H$_\alpha$ and Ba\,II
6496~\AA\ line profiles. (c)~the Na\,I D$_2$D$_1$ and Ba\,II
5854~\AA\ line profiles. } \label{figure7:Goranskij_n}
\end{figure*}

The whole spectrum obtained on February 24, 2015 is shown in
Fig.~\ref{figure6:Goranskij_n}, where the strongest lines are
identified. Figure~\ref{figure7:Goranskij_n} shows the fragments of
this spectrum with identification of the weaker lines. The continuum
of the cool star predominates in the spectrum. From the collection of
spectra by \cite{Jacoby84:Goranskij_n}, the star HD\,1069 (K2\,I)
approximates the energy distribution of the LRN well. The H$\alpha$
line is asymmetric, fully in emission with the intensity maximum at a
velocity of $300$~km\,s$^{-1}$, equivalent width ${\rm EW}=28$~\AA,
instrumental-profile corrected \mbox{${\rm FWHM}=535$~km\,s$^{-1}$}.
Note that the heliocentric velocity of the galaxy M\,101 equals to
$241\pm2$~km\,s$^{-1}$ (from the NED database). The full width at
zero intensity of the line equals to ${\rm FWZI}=1370$~km\,s$^{-1}$.
The Ba\,II $\lambda$6496~\AA\ line, the Ba\,II $\lambda$6136,
6142~\AA\ blend, and the Na\,I~D$_2$D$_1$ $\lambda$5890, 5896~\AA\ blend
have P\,Cygni profiles, the maximum of the emission
intensity in these profiles is at $330$~km\,s$^{-1}$. The
minima of absorption components of these lines are located at
$-260$~km\,s$^{-1}$, and the absorption spread till
$-620$~km\,s$^{-1}$. The lines of H$\beta$, Ba\,II $\lambda$4709, 4957,
5874~\AA, Mg\,I $\lambda$5167, 5173~\AA, and numerous Fe\,I, Ti\,I,
and Cr\,I lines are observed in the absorption. Radial velocities of
these lines are similar to the velocities in absorption
components of P\,Cygni profiles of the strong lines. The set
of absorption lines of chemical elements duplicates in detail the set
of lines for the red nova  V838\,Mon in the January 2002 outburst,
when its spectrum was classified as K0\,I \citep{Goray02:Goranskij_n}.
The significant difference of the LRN spectrum was the absence of the
Li\,I $\lambda$6708~\AA\ line, which was very strong in the V838\,Mon
spectrum and having a P\,Cygni profile.

\begin{figure*}
\includegraphics[scale=0.65]{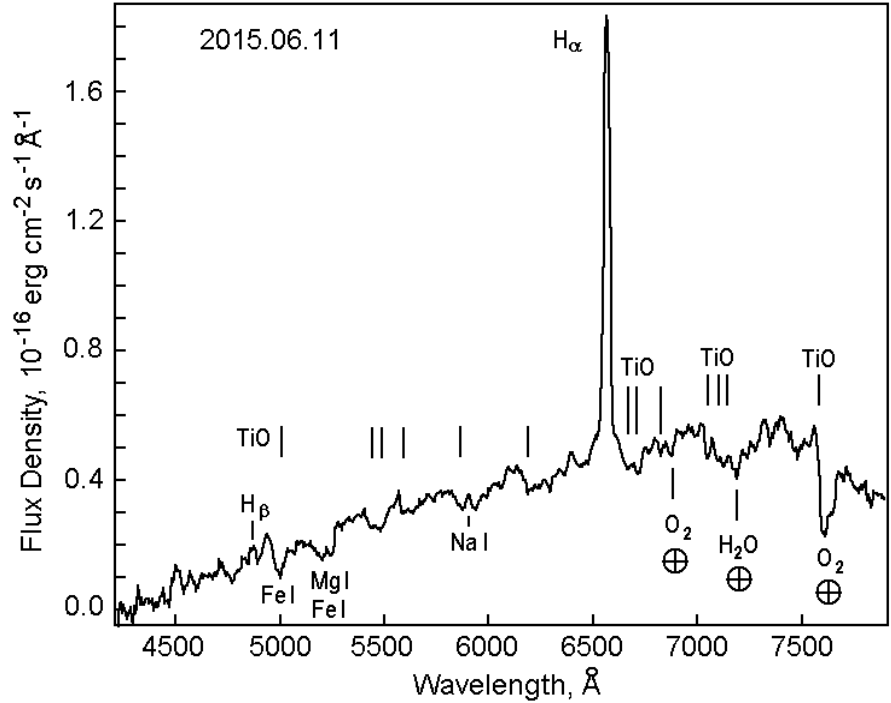}
\caption{Spectrum of the LRN in M\,101 obtained
on the BTA telescope with the SCORPIO reducer on June 11, 2015. The
original spectrum with a small $S/N$ ratio is averaged with the
window 14~\AA\ which corresponds to the spectral resolution. The symbol
$\bigoplus$ denotes the molecular bands of the terrestrial origin. }
\label{figure8:Goranskij_n}
\end{figure*}

From the observations carried out in February, the velocity of the
envelope expansion of the LRN in M\,101 at the peak of the second
outburst was $500$--$540$ km\,s$^{-1}$. By comparison, the velocity of
the envelope expansion of V838\,Mon at the outburst peak was only
$150$~km\,s$^{-1}$, which was 3.5 times less.

Cross-correlation of the LRN/M101 normalized spectrum of February 24,
2015 with the normalized spectra of the supergiants by
\cite{Jacoby84:Goranskij_n} determines the LRN spectrum in a wider
range as \mbox{K0\,I--K5\,I} (from spectral lines). The strength of
the absorption lines in the LRN/M\,101 spectrum 4--6 times exceeds
the strength of the lines of normal stars. Thus, the depressions
of the continuum in the ranges \mbox{$\lambda$5000--5500~\AA} and
\mbox{6100--6400~\AA} are due to line blanketing.

Figure~\ref{figure8:Goranskij_n} shows the red nova spectrum obtained
on June 11, 2015. This spectrum has resolution three times lower than
that one obtained on February and, therefore, the narrow absorption
lines are seen not so clearly. Energy distribution in the spectrum
shifted toward the long-wave region. It can be best approximated with
the distribution of the star  HD\,13136 (M2\,Ib), although, there are
no later spectral type stars among the class I luminosity objects for
comparison in the collection by \citet{Jacoby84:Goranskij_n}. This is an
\mbox{M1\,I--M4\,I} type star by the color index
$B-V=1\fm83\pm0\fm10$. The dominant  H$\alpha$ emission with an
equivalent width of  ${\rm EW}=107$~\AA\ is more symmetric than that
in the February spectrum. Its instrumental-profile corrected
halfwidth is ${\rm FWHM}=900$~km\,s$^{-1}$. One can see weak emission
wings in this line which determine the full width of ${\rm
FWZI}=189$~\AA\ ($8600$~km\,s$^{-1}$). These wings are probably
formed due to Thomson scattering. There may be a weak emission in
H$\beta$ and \mbox{Na\,I~D$_2$D$_1$}.

The TiO bands can be noticed in the absorption spectrum, the
strongest of which have heads at wavelengths of $\lambda$4955, 5450,
5498, 5597, 6159, 6659, 6715, 6817, 7054, 7090, 7126~\AA, and the
strongest one---at $\lambda$7589~\AA. Atomic absorptions can be
hardly seen due to the low resolution, only the Mg\,I
$\lambda$5167--5183~\AA\ triplet  and some  Fe\,I blends can be
identified.

Spectral observations, as well as the photometry, show the energy
distribution shift towards the long-wave region. These are
low-excitation spectra in which the atomic lines predominate and then
the molecular bands appear when the brightness declines. Such
development is characteristic of red novae.

\section{DISCUSSION OF RESULTS}

The LRN in M\,101 proves to be a massive young star, the outburst of
which took place in the spiral arm of the M\,101 galaxy in the region
of the hot OB-supergiant association. It is not associated with
nebulae. Red novae belong to different types of galaxy populations
and show different amplitudes, outburst durations, and shapes of
light curves. PTF10fqs emerged in the spiral arm of the M\,99 galaxy
and was associated with the OB-supergiant
association~\citep{Kasli11:Goranskij_n} as well as the LRN in M\,101.
The other red nova PTF10acbp \cite{Kasli10:Goranskij_n} in the spiral
galaxy UGC\,11973 has exploded on the outskirts of this galaxy and
obviously belongs to the young population of the disk. The red nova
OT\,2006-1~\citep{Kasli10:Goranskij_n} in the galaxy  M\,85,
classified as S0, has appeared on the edge of the galaxy and is not
evidently associated either with the H\,II region or with any other
star formation region \citep{Ofek08:Goranskij_n}. The red nova
V838\,Mon is associated with the population~I of our Galaxy and with
the young cluster of B stars \citep{Afsar07:Goranskij_n}. V1006/7 in
M\,31 and the Galactic red novae V4332\,Sgr and V1309\,Sco are the
objects of a galactic bulge or a thick disk. In the V4332\,Sgr system
there is an evolved star, a red giant, the radiation from which is
detected in energy distribution of the outburst progenitor and also
in the spectrum after the outburst \citep{Bars14:Goranskij_n}. The
last three objects are obviously old star systems which got through
the long-term evolution process.

The duration of red nova outbursts varies in the range of 58 days for
\mbox{M31N\,2015-01a} (in the $V$ filter) to 135 days for
PTF10fqs/M\,99 (in the $R$ filter), if the duration is estimated as
a time when the star resides over the level 3$^{\rm m}$ below the brightness
maximum. Red nova outbursts have similar light curves: slow
brightness weakening immediately after the maximum (``flat maximum''
or ``plateau'') which ends in the steep fall, although there are some
exceptions. The LRN in M\,101 exceeds all other red novae in the
outburst duration. If we estimate the second outburst duration by
this criterion, we obtain 154~days (in the $V$ filter). In case we
count from the first outburst in November 2014, which was brighter at
the maximum, we obtain the outburst duration longer than 154~days.
There are some other special features of the light curve of the LRN
in M\,101 which become apparent when compared to other red novae. In
the second outburst in February 2015, neither flat maximum nor steep
fall were observed. Moreover, there was the pause of the brightness
decline for 30~days. It is interesting that in the V1309\,Sco red
nova there were two similar stops of the brightness variation (see
fragments 9 and 10 in the light curve in \hbox{Fig.~1 and 6} by
\cite{Tylen11:Goranskij_n}). These stops in the brightness decline
can also be associated with the outgoing of the weak shockwaves to
the envelope surface. It should be noted that the shape of the
V1309\,Sco light curve also differed from those of other red novae with
the absence of the flat maximum, its brightness decline after the
outburst was gradual and with the following slowdown similar to that
of the LRN in M\,101.

If we take the minimum brightness observed in 1993 (DSS, POSS\,II) as
a reference point for the LRN in M\,101, then the outburst amplitude
will be equal to $5\fm6$, which is the lowest amplitude measured for
a red nova. Still, in absolute magnitude at the brightness first
maximum, $M_V=-12\fm75$, it is probably the most luminous red nova.
The star OT\,2006-1 in the galaxy  M\,85 reached the same absolute
magnitude but in the red spectral domain, $M_R=-12\fm7$
\citep{Pastor07:Goranskij_n}, the authors classified it as SN\,IIp.

The characteristic feature of the light curve of the LRN in M\,101 is
the gradual increase of brightness before the outburst, which was at
least $2\fm2$. As can be seen from the observations of the red nova
V1309\,Sco \citep{Tylen11:Goranskij_n}, such an increase in brightness
takes place when the common envelope is forming before star merging
in a contact system; thus, the event of the LRN in M\,101 in 2015
most probably is a merger. In the case of V1309\,Sco, the gradual
brightness increase stopped a year before the maximum with the abrupt
weakening of brightness by $1\fm$ In the case of the LRN/M\,101, the
brightness weakening was not detected between the last observation at
the brightness increase stage in July 2014 (LBT,
\citet{Gerke15:Goranskij_n}) and the first observation at the outburst
peak on November 10, 2014 (PTF, \citet{Cao15:Goranskij_n}). Such
brightness weakening can be expected when the common envelope is
expanding in the adiabatic-like mode after the energy burst and the
internal impulse at the merging of two stars' cores. However, we can
see the result of such an expansion in the second outburst as a
considerable reddening of the color indices. Under the assumption
that the common envelope had already formed by July 2014,  we
estimated the variations of its radius using the Stefan--Boltzmann
law and the photometric parameters of the star. The envelope photosphere
has increased by eight times approximately from $400~R_\odot$ in July 2014 to $3300~R_\odot$
in February 2015. In June 2015, it grew up to
$4700~R_\odot$. These estimates can be inaccurate because of the
complex structure of the outburst remnant, however, they prove that
the expansion of the envelope photosphere happened in the
period between July 2014 and February 2015. The velocities determined from
the absorption lines and components in the P\,Cygni profiles
($500$--$540$~km\,s$^{-1}$) do not correspond to such envelope radius
variations estimated with the Stefan--Boltzmann method, and 2--3
times exceed these estimates. The similar effect was also observed in
V838\,Mon.

Taking into account that the expansion velocities of the LRN in
M\,101 based on the absorption lines and components are three times
more than the velocities of the envelope expansion of V838\,Mon
and V1309\,Sco at this stage, the duration of the envelope expansion
can be significantly shorter, only several months. As the outburst
developed at such great rates, it could be assumed that the shockwave
formed after the stars' cores merging and the energy impulse from the
inside, and the first outburst in November 2014 was associated with
emerging of this shockwave. It was most probable that thereby some
line absorbing layer separated from the envelope and moved at a
higher velocity than the photosphere. It is due to the absorption of light
in this additional layer both in the LRN in M\,101 and in V838\,Mon, the
absorption spectrum lines are several times stronger than the lines
of normal stars of the same spectral type.

After emerging of the shockwave, in December 2014 and January 2015,
the envelope expanded with the decrease of energy output, which
resulted in the brightness decrease at least by $2\fm4$. We assume
that this brightness decline is similar to that brightness decrease
by $1^{\rm m}$ which we observed a year before the maximum in
V1309\,Sco. It is not excluded that the shockwave formation in the
LRN in M\,101 is associated with the high mass of this system. The
second outburst in February 2015 is due to an arrival of the
outburst thermal energy to the surface of the expanding envelope.
This scenario can be confirmed or disproved by the dynamic model
calculations or by the analysis of the archival data obtained in the
period between July 2014 and February 2015.

\section{CONCLUSIONS}

The LRN in M\,101 has emerged in the spiral galaxy in the region of
the OB-star association and is a massive system, the bright component
of which leaves the  main sequence and shifts toward the red region
in the ``color-magnitude'' diagram. Over 11 years prior to the
outburst, the brightness of the system gradually increased by
$2\fm2$. As the brightness increase before the outburst takes place
when the components form a common envelope and spiral towards each other, we
identify this event with a merger of the components in a massive
system.

The star had an unusual light curve with two maxima. It was
discovered in the repeated outburst that followed in three months after
the first maximum, in which it has reached the visual absolute magnitude
\mbox{$M_V = -12\fm75$}. The star envelope expanded prior to the
repeated outburst increasing its radius by a factor of 8. In the
maximum of the repeated outburst, the star spectrum was approximately
classified as K2\,I. We detected the H$\alpha$ emission, strong
Ba\,II and Na\,I lines with the P\,Cygni profiles, and the extremely
strong absorption in the Fe\,I, Ti\,I, Cr\,I, and Mg\,I metal lines.
Formation of such a spectrum is obviously associated with the
ejection of the absorbing layer at the shockwave outgoing during
the first outburst. The rates of the absorbing layer expansion
are\linebreak $500$--$540$~km\,s$^{-1}$. Then within four months after
the maximum at the same time with the temperature decrease, the spectral type was
changing to about M2\,I. The weak TiO molecular bands emerged. The
spectral development is typical for the red novae and leaves no doubt
about the classification.

The outburst of the LRN in M\,101 is a record of the duration among
the known red novae (>153~days), of the absolute visual
magnitude at the maximum ($M_V = -12\fm75$), and of the outburst
amplitude ($5\fm6 V$) which is the smallest of those ones observed in
red novae.

\section{Acknowledgments}

In the paper, we used the following databases: Sloan Digital Sky
Survey, NASA/IPAC Extragalactic Database (NED), Vienna Atomic Line
Database (VALD), NIST Atomic Spectra Database, SuperCOSMOS Sky
Survey, and image web hostings AstroBin and Flickr. The spectral and
photometric observations carried out at SAO~RAS, their reduction and
analysis were financed with the grant of the Russian Science
Foundation No.~14-50-00043. The operation of the Russian 6-m BTA
telescope is carried out with the financial support of the Ministry
of Education and Science of the Russian Federation (agreement
No.~14.619.21.0004, project ID RFMEFI61914X0004). In our work, we
used the instruments created with the support from the
M.~V.~Lomonosov Moscow State University Development Program. VPG,
EAB, and AFV are grateful to the Russian Foundation for Basic
Research for the financial support in the present research with the
grant 14--02--00759. ASM wishes to thank the President of the Russian
Federation for the financial support with the grant MK-1699.2014.2.

\bibliographystyle{mnras}
\bibliography{M101_n_en} 

\bsp	
\label{lastpage}
\end{document}